\documentclass[a4paper,conference]{IEEEtran}
\usepackage{cite}

\usepackage{graphicx}
\usepackage[cmex10]{amsmath}
\usepackage{amssymb}
\usepackage{wasysym}

\usepackage{booktabs}
\usepackage[tight,footnotesize]{subfigure}

\usepackage{hyperref}
\usepackage{url}

\usepackage{flushend}

\usepackage{fancyhdr}
\begin{document}
%
\title{Shoreline and Bathymetry Approximation in Mesh Generation for Tidal Renewable Simulations}
\author{\IEEEauthorblockN{Alexandros Avdis\IEEEauthorrefmark{1}\IEEEauthorrefmark{2},
Christian T. Jacobs\IEEEauthorrefmark{1},
Jon Hill\IEEEauthorrefmark{3},
Matthew D. Piggott\IEEEauthorrefmark{1} and
Gerard J. Gorman\IEEEauthorrefmark{1}}
\IEEEauthorblockA{\IEEEauthorrefmark{1}Department of Earth Science and Engineering, Imperial College London, UK}
\IEEEauthorblockA{\IEEEauthorrefmark{2}E--mail: \tt a.avdis@imperial.ac.uk}
\IEEEauthorblockA{\IEEEauthorrefmark{3}Environment Department, University of York, UK}\\
\IEEEauthorblockA{\textbf{Pre--print of conference publication accepted in the Proceedings of}}
\IEEEauthorblockA{\textbf{11th European Wave \& Tidal Energy Conference (\href{http://www.ewtec.org/proceedings/}{EWTEC2015})}}}


\maketitle

\begin{abstract}
Due to the fractal nature of the domain geometry in geophysical flow simulations, a completely accurate
description of the domain in terms of a computational mesh is frequently deemed infeasible.
Shoreline and bathymetry simplification methods are used to remove small scale details in the
geometry, particularly in areas away from the region of interest. To that end, a novel method for shoreline
and bathymetry simplification is presented. Existing shoreline simplification
methods typically remove points if the resultant geometry satisfies particular geometric criteria.
Bathymetry is usually simplified using traditional filtering techniques, that
remove unwanted Fourier modes. Principal Component Analysis (PCA) has been used in other fields to isolate
small--scale structures from larger scale coherent features in a robust way, underpinned by a
rigorous but simple mathematical framework. Here we present a method based on principal component
analysis aimed towards simplification of shorelines and bathymetry. We present the algorithm in
detail and show simplified shorelines and bathymetry in the wider region around the North Sea.
Finally, the methods are used in the context of unstructured mesh generation aimed at tidal
resource assessment simulations in the coastal regions around the UK.
\end{abstract}


%
\IEEEpeerreviewmaketitle

\section{Introduction}
Ocean and coastal models are routinely used to assess
the tidal energy potential of a site as well as any potential environmental impacts
due to the presence of energy extraction devices (e.g. \cite{Hasegawa2011, Funke2014-vk, Martin-Short2015}).
Shoreline contour databases are used to define the simulation domain
and create a computational mesh. However, while smaller scale structures in the shoreline
may be relevant in the region of primary interest to the
given study, a coarser representation is often sought in other regions.
For example, small mesh--cells are required in the vicinity of tidal turbines
and their wakes, for accurate calculation of power generation or scour patterns.
In addition, the accurate prediction of currents in the complex geometries typical
of tidal energy sites requires higher mesh resolution near the often intricate shoreline,
including small islands, and near areas of steep bathymetry.
However, in areas away from the region of
interest the smaller scale geometries must be removed in order to alleviate the otherwise
stringent requirements on mesh resolution, and therefore reduce computational costs.
In addition, as discussed in \cite{gorman:2007} and \cite{gorman:2008} an automated shoreline simplification and
parameterisation method is paramount to ensure the (reproducible) generation of high quality meshes with minimal
user intervention. To facilitate the use of multi-scale simulation methods we only
consider unstructured meshes in this work.
\par
Existing methods of \emph{simplification} (or \emph{smoothing}) are typically based on geometric
criteria, applied to all points in a piece--wise linear contour and involve the modification
of point coordinates and/or the removal of points. The method proposed in \cite{Ramer:1972} and
\cite{douglas_peucker:1973} is perhaps the most widely used example of this type
of contour simplification. In fact, the simplified land masses in the Global, Self--consistent,
 Hierarchical, High--resolution Shoreline (GSHHS)
data sets in \cite{wessel_smith:1996} have been created using the algorithm
proposed by Douglas and Peucker \cite{douglas_peucker:1973}. However, in many cases
the resulting geometry can be unsuitable for mesh generation. In particular, the lack
of smoothness in the simplified geometry can force poor quality elements from the mesh generator.
In addition, consistency between bathymetry and the shoreline contour is sometimes necessary, where
the bathymetric map gives a value of approximately the correct value at the given contour. Thus
a shoreline is extracted from a bathymetric map and subsequently simplified. In terms of bathymetry
simplification, filtering of higher frequency components using Fourier analysis is the obvious method
for simplification.
\par
Here we present algorithms targeted towards simplification of shorelines and
bathymetry based upon \emph{Principal Component Analysis (PCA)}. While PCA has been applied
in many areas including raster composition from measurements, raster
analysis \cite{Demsar_et_al:2013} and beach
morphodynamics studies \cite{ruessink_et_al:2004, medina_et_al:1992, winant_et_al:1975},
the application of PCA towards shoreline and bathymetry simplification is novel.
The theoretical
framework underpinning the proposed method creates a very robust and efficient geometry simplification
method that simplifies the domain geometry in a reproducible
way. We demonstrate methods for smoothing realistic shorelines
and bathymetry and showcase the utility of such methods in mesh generation, geared towards
assessment of tidal renewable energy in coastal regions around the UK.

\section{Shoreline and Bathymetry Simplification Algorithms}
Principal component analysis was developed as a multivariate analysis method through the
work of  Galton\cite{galton:1889},  Pearson \cite{pearson:1901}
and  Hotelling \cite{hotelling:1933, hotelling:1936}. PCA is used to identify
dominant structures or patterns in data, for example in image analysis and compression
\cite{rosenfeld_kak:1982} and turbulence structure analysis \cite{holmes_lumley_berkooz:1996}.
The aim of PCA is to reduce the dimensionality of the given data, such that the maximum
possible variance of the input data is retained. This is achieved by projecting the
data onto a set of uncorrelated basis functions. The projections are termed \emph{principal
components}. A more complete description of the theoretical background of PCA is beyond the
scope of this paper, but it can be found in \cite{jolliffe_pca_book:2002} and \cite{wold_et_al:1987}.
We here present a procedural view--point of PCA for completeness.
Briefly, let $u(x)$ be an observed variable of a given system, and we
repeat $M$ experiments using that system. We denote the outcome of
the experiments as $u_i(x)$, $i \in {1,2,\ldots,M}$.
The discrete form of PCA is of particular interest here,
where $x$ denotes discrete points where data is provided. 
Each observation $u_i(x)$ is thus structured as a vector, storing the value at the $N$ discrete points:
\begin{equation}
\mathbf{u}_i = \left[u_i(x_1), u_i(x_2), \ldots, u_i(x_n), \ldots, u_i(x_N) \right]^T.
\label{eqn:PCAdiscreteSamples}
\end{equation}
The average is removed from each observation and only
the vectors of fluctuations $\tilde{\mathbf{u}}_i$ are subsequently used.
For notational convenience let $\mathbf{S}$ denote the $N \times M$ matrix of all $M$ observations:
\begin{equation}
\mathbf{S} = \left[ \tilde{\mathbf{u}}_1, \tilde{\mathbf{u}}_2, \ldots, \tilde{\mathbf{u}}_M \right].
\label{eqn:PCASamples}
\end{equation}
PCA produces a decomposition of the data in $\mathbf{S}$ as a linear combination of a
set of $N$ modes:
\begin{equation}
\mathbf{S} = \mathbf{A} \mathbf{\Phi}^T,
\label{eqn:PCA_reconstruction_matrixForm}
\end{equation}
where $\mathbf{\Phi}$
is an $M \times N$ matrix of the PCA modes. The eigen--vectors ordered
into the $N \times N$ matrix $\mathbf{A}$ are obtained from:
\begin{equation}
\mathbf{C} = \mathbf{A} \mathbf{\Lambda} \mathbf{A}^T,
\label{PCA_eigenproblem}
\end{equation}
where $\mathbf{\Lambda}$ is the diagonal matrix of eigenvalues and $\mathbf{C}$ is the $N \times N$ covariance matrix:
\begin{equation}
\mathbf{C} = \frac{1}{M} \mathbf{S} \mathbf{S}^T.
\label{eqn:PCA_covarianceMatrix}
\end{equation}
From an algorithmic point--of--view principal component analysis can be broadly described by
the following main steps:
\begin{enumerate}
\item Collect $M$ observations of the system and order the data from each observation as an
$N$--dimensional vector, removing the average from each observation. Then assemble the matrix
$\mathbf{S}$.
\item Construct the covariance matrix $\mathbf{C}$ using equation \eqref{eqn:PCA_covarianceMatrix}.
\item Orthogonalise the covariance matrix, as described in equation \eqref{PCA_eigenproblem}.
\item Calculate the PCA modes from an inversion of equation \eqref{eqn:PCA_reconstruction_matrixForm}: 
\begin{equation}
\mathbf{\Phi} = \mathbf{S}^T \mathbf{A}.
\label{eqn:PCA_modesCalculation}
\end{equation}
\end{enumerate}
A key property of PCA is that the first mode represents the most energetic ``structure'' in the input data,
followed by the second mode, and so forth. Thus synthesis from the dominant PCA modes
will preserve the most important structures while affecting geometry simplification.
\par
In sections \ref{ssect:ShorelineSimplification}
and \ref{ssect:BathymetrySimplification} we discuss how we have adapted the
steps outlined above towards shoreline and bathymetry simplification.
\subsection{Shoreline simplification}
\label{ssect:ShorelineSimplification}
\begin{figure}[ht!]
\centering
\includegraphics[width=\columnwidth]{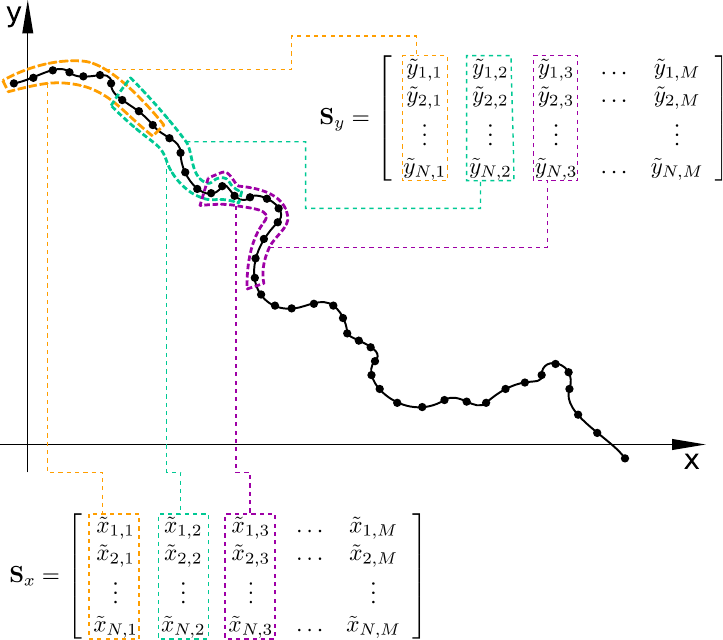}
\caption{Diagram of line partitioning. The black circular markers denote the points of the piece--wise linear contour.
The colour dash lines show how partitions can be constructed from the lines, with a partial overlap. The
$x$--coordinates of the points are used to construct matrix $\mathbf{S}_x$ and $y$--coordinates are used
in matrix $\mathbf{S}_y$. As shown, each column of the matrices is constructed by listing the corresponding
partition point coordinates.}
\label{fig:linePartitioning}
\end{figure}
\begin{figure}[ht!]
\centering
\includegraphics[width=\columnwidth]{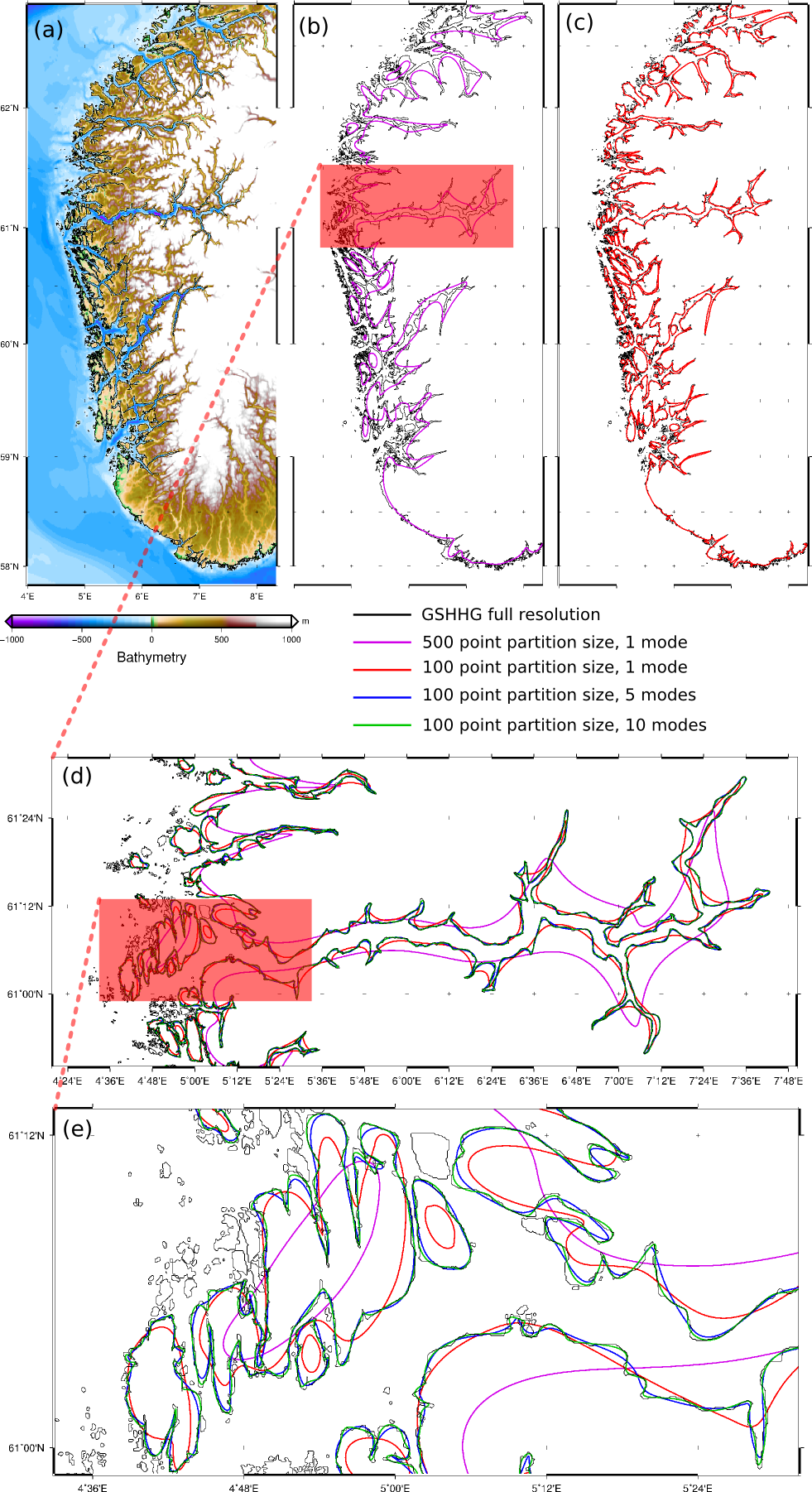}
\caption{Simplifications of the GSHHG full resolution data--set over the West Norwegian coast, with detailed
focus on Sognefjorden. (a) West Norwegian coast with the GSHHG full resolution shoreline and the GEBCO
$30^{\prime\prime}$ elevation map. (b) The GSHHG full resolution shoreline and the reconstruction from
the first mode of PCA simplification using $500$ point partitions. (c) The GSHHG full resolution
shoreline and the reconstruction from the first mode of PCA simplification using $100$ point partitions.
(d) Sognefjorden and surrounding shorelines from GSHHG full resolution and various PCA reconstructions.
(e) Sognefjorden entrance and islands from GSHHG full resolution and various PCA reconstructions. Islands
too small to construct a partition are not reproduced in the reconstruction output, affecting small island
filtering.}
\label{fig:shorelineReconstructionMaps}
\end{figure}
Shoreline data typically consists of a number of piece--wise linear line contours, each defined as a list of points.
Our proposed method of shoreline simplification consists of applying a PCA algorithm to each contour
in turn. Figure \ref{fig:linePartitioning} shows how the decomposition of a line contour has been
implemented. The $M$ samples in equation \eqref{eqn:PCASamples} are formed by partitioning the contour,
and each column of the $\mathbf{S}$ array corresponds to a partition. Figure \ref{fig:linePartitioning} also shows that
the $x$--coordinates are treated separately to the $y$--coordinates, with separate $\mathbf{S}$ 
matrices assembled for each. Two separate eigen--problems are
solved for $\mathbf{C}_x = \dfrac{1}{M}\mathbf{S}_x \mathbf{S}^T_x$ and
$\mathbf{C}_y = \dfrac{1}{M}\mathbf{S}_y \mathbf{S}^T_y$ covariance arrays. The partitions are next
reconstructed from the chosen dominant modes. As suggested in figure \ref{fig:linePartitioning},
successive partitions are allowed to overlap. The final step is the re--assembly of the contour from the
reconstructed partitions and is done on a point--by--point basis: The coordinates of points where partitions overlap are
calculated by averaging the coordinate values of the given point, across the overlapping reconstructed partitions.
Where partitions do not overlap, the point coordinates are used directly in the corresponding assembled contour point.
It has been found that best smoothing results are obtained when the number of partitions is
equal to the number of points in a contour. In this way each partition corresponds to a point in the contour.
However, when the number of points in a contour is less
than the specified number of points per partition, the contour is not considered for decomposition
and not reproduced in the output. This has been found to be a very effective way of filtering out
small islands, as shown in figure \ref{fig:shorelineReconstructionMaps}.
\subsection{Bathymetry simplification}
\label{ssect:BathymetrySimplification}
\begin{figure}[ht!]
\centering
\includegraphics[width=\columnwidth]{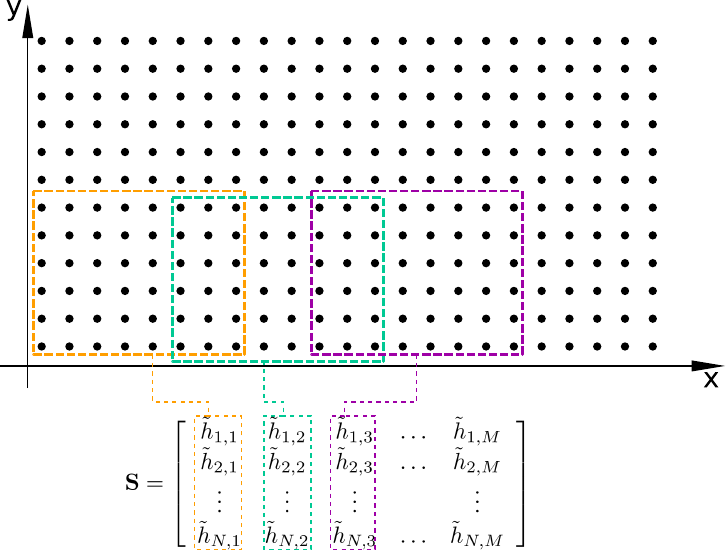}
\caption{Diagram of raster partitioning. The black circular markers denote the points of the raster grid.
The colour dash lines show how the partially overlapping partitions can be constructed. As shown, each
column of matrix $\mathbf{S}$ corresponds to a partition.}
\label{fig:rasterPartitioning}
\end{figure}
The implementation of smoothing for bathymetry assumes the data to be given as a ``raster''
where data points are laid out as a rectangular
grid, such that the points are aligned along the coordinate directions and each point
can be identified by a pair of integer indices. Figure \ref{fig:rasterPartitioning} illustrates the
topological structure of the data-points and shows that, as in shoreline simplification,
the $M$ samples in \eqref{eqn:PCASamples} are obtained by partitioning the data. Figure
\ref{fig:rasterPartitioning} also shows that partitions are allowed to partially overlap. As with
shoreline reconstruction, best smoothing results are obtained when the number of partitions is
equal to the number of points of the input bathymetry. Once the partitions are reconstructed from the dominant modes, the value
at any point is calculated as the average of the reconstructions of all partitions overlapping
at the given point.

\section{Results}
\subsection{Shoreline simplification on the Western Norwegian coast}
\begin{figure*}[ht!]
\centering
\includegraphics[width=\textwidth]{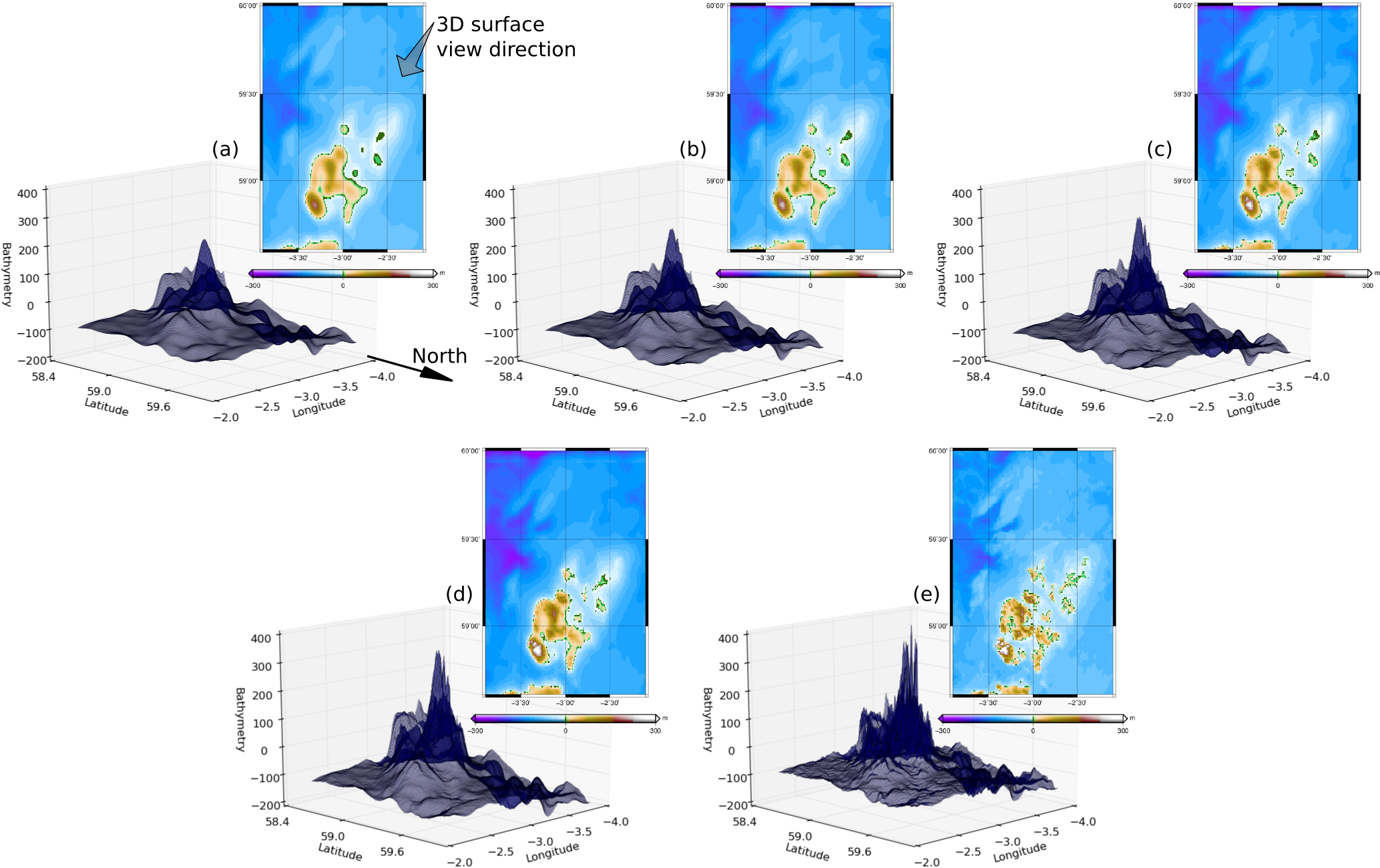}
\caption{Various raster approximations of the bathymetry/topography around the Orkney Islands,
extracted from the GEBCO $30^{\prime\prime}$ 2014 data--set. The region is identified by a semi--transparent
red rectangle in figure \ref{fig:simulationDomainMaps}(a). Data is visualised as maps (smaller inserts) and
semi--transparent, three--dimensional surface plots. The viewing angle of the surface plots is indicated by the arrows
in panel (a), showing the reconstruction using only the first PCA mode. (b) Raster reconstruction using
the first $16$ modes. (c) Raster reconstruction using the first $32$ modes. (d) Raster reconstruction using
the first $48$ modes. (e) GEBCO $30^{\prime\prime}$ 2014 raster data.}
\label{fig:bathymetryReconstructionMaps}
\end{figure*}
\label{ssect:ShorelineSimplificationResults}
The Western Norwegian coast is used here as an exemplar of the effectiveness of the shoreline simplification
method outlined in section \ref{ssect:ShorelineSimplification}. This area was chosen due to the geometrical complexity of its
shorelines, with many small islands, but also because a relatively coarse approximation to this shoreline
is used in unstructured meshes aimed at tidal flow simulations discussed in section \ref{ssect:meshGeneration} below. 
\par
Figure \ref{fig:shorelineReconstructionMaps} identifies the region and also shows the results of various
reconstructions superimposed on the ``full resolution'' Global, Self--consistent, Hierarchical, High--resolution
Geography (GSHHG) data. Panels (b) to (e) show that many
small islands have been removed. As discussed in section \ref{ssect:ShorelineSimplification} contours
whose point--count is smaller than the segment size cannot be decomposed and are not reproduced in
the reconstruction. However, islands composed of too few points are likely to be of little utility
to a shoreline approximation based on a larger segment size, where small scale islands are desired
to be removed in the first place. Figures \ref{fig:shorelineReconstructionMaps}(b) and
\ref{fig:shorelineReconstructionMaps}(c) show that increasing the segment size imparts greater
smoothing on the shoreline reconstruction. The reconstruction in figure \ref{fig:shorelineReconstructionMaps}(b)
was obtained using a segment size of $500$ points and only the first PCA mode. Only the very large scale features
are captured. The first PCA mode is also solely used in the reconstruction of figure
\ref{fig:shorelineReconstructionMaps}(c), but with a smaller segment size, of $100$ points. As a result,
the reconstruction in panel (c) captures a lot more detail compared with that in panel (b).

\subsection{Bathymetry simplification on the Orkney Islands.}
\label{ssect:BathymetrySimplificationResults}
The PCA--based simplification method described in section \ref{ssect:BathymetrySimplification} was used
to generate the plots in panels (a)--(d) of figure \ref{fig:bathymetryReconstructionMaps}. The bathymetry,
shown in figure \ref{fig:bathymetryReconstructionMaps}(e), is an excerpt from the GEBCO $30^{\prime\prime}$
2014 data--set, over the Orkney Islands. The region is indicated in figure \ref{fig:simulationDomainMaps}(a).
The partition size was $8 \times 8$ points, resulting in a total of $64$ PCA modes. The reconstructions
shown in figures \ref{fig:bathymetryReconstructionMaps}(a)--(d) use successively more PCA modes: Just the
first mode in panel (a), $16$, $32$ and $48$ modes in panels (b), (c) and (d) respectively. $64$ modes
amount to reproducing the input data, shown in panel (e). The first mode captures the most important features
of the bathymetry with successive modes adding more details to the reconstruction. The highest peak in the region
is in the Isle of Hoy, clearly visible as a prominent peak in all reconstructions, along with the Scottish mainland
coast (just behind the Hoy peak in the surface plots) and the bathymetry troughs north--west of the Orkney Islands.
\subsection{Shoreline and bathymetry simplification in meshing for tidal flow simulations of UK coastal regions.}
\label{ssect:meshGeneration}
\begin{figure*}[ht!]
\centering
\includegraphics[width=\textwidth]{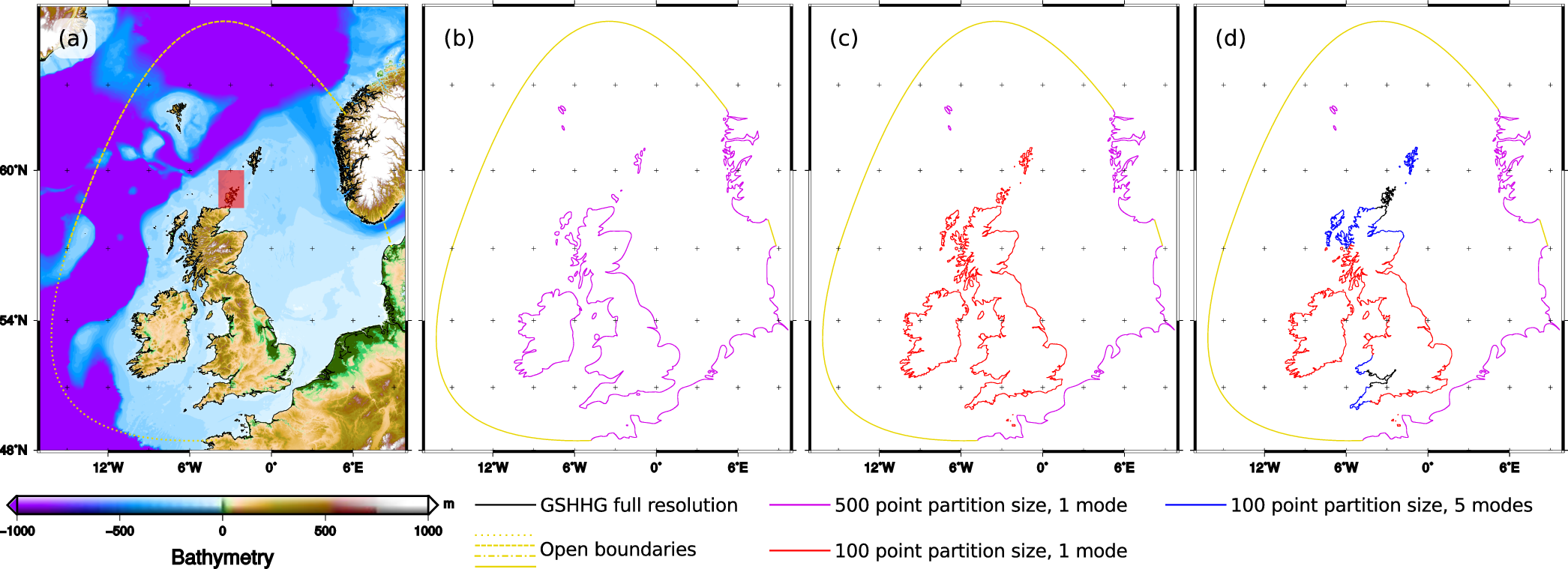}
\caption{Shorelines, domain boundaries and bathymetry map of the simulation domain. The violet, red and
blue lines denote reconstructions from a PCA simplification of the GSHHG full resolution shoreline. The colour--line
convention is the same as in figure \ref{fig:shorelineReconstructionMaps}. (a) GSHHG full resolution
shorelines (black lines) and open boundaries (yellow lines) superimposed on the GEBCO $30^{\prime\prime}$ 2014
bathymetry. The semi--transparent red rectangle indicates the region used in the raster simplification results in
figure \ref{fig:bathymetryReconstructionMaps}. (b) Domain boundaries for the ``coarse'' mesh. 
(c) Domain boundaries for the ``intermediate'' mesh. (d) Domain boundaries for the ``fine'' mesh. All panels
drawn in Mercator projection.}
\label{fig:simulationDomainMaps}
\end{figure*}

We here showcase the shoreline and bathymetry simplification in a simulation of tidal flow
around the United Kingdom. As figure \ref{fig:simulationDomainMaps} shows, the simulation
domain includes the North Sea, English Channel, Saint George's Channel, Irish Sea and part
of the Northern Atlantic. However, the focus of this study is two sites of particularly high
potential for renewable energy generation from tides: The Orkney Islands and the
Severn Estuary \cite{Martin-Short2015, willis_et_al:2010}.
\par
The GSHHG ``full resolution'' \cite{wessel_smith:1996} data--set, at a resolution of $200$ metres,
was used as the source of the shorelines shown in figure \ref{fig:simulationDomainMaps}(a). Panels
(b), (c) and (d) in figure \ref{fig:simulationDomainMaps}
show how shoreline data and shoreline reconstructions have been combined to define the domain boundaries for a
``coarse'' (panel b), ``intermediate'' (panel c) and ``fine'' (panel d) mesh.
The open boundaries shown by yellow lines in figure \ref{fig:simulationDomainMaps} are constructed by
combining lines of constant bearing (\emph{loxodromes}). The dashed yellow line in figure \ref{fig:simulationDomainMaps}(a)
was obtained by combining two loxodromes: The first loxodrome is drawn from point
$15^\circ W$, $57^\circ N$, at an angle to the North pointing axis (\emph{bearing}) of $20^\circ W$, up to $70^\circ N$.
The second loxodrome is drawn from point $5^\circ:27^\prime E$, $62^\circ N$, at a bearing of $0^\circ$, up to $70^\circ N$.
The two loxodromes are combined linearly, such that the loxodrome starting points are the end points of
the dashed line in figure \ref{fig:simulationDomainMaps}(a). The dotted yellow line is a combination
of the loxodrome from $4^\circ W$, $48^\circ:30^\prime N$ at bearing $110^\circ W$ up to $25^\circ E$ with the loxodrome
from $5^\circ:27^\prime E$, $62^\circ N$ at bearing $185^\circ W$ up to $55^\circ N$. The dashed--dotted line across
Skagerrak is a loxodrome from $8^\circ E$,$58^\circ:5^\prime N$ at bearing $30^\circ W$ up to $57^\circ N$.
The lines are then trimmed at the intersections with the shorelines to close the domain.
\par
\begin{figure*}[ht!]
\centering
\includegraphics[width=\textwidth]{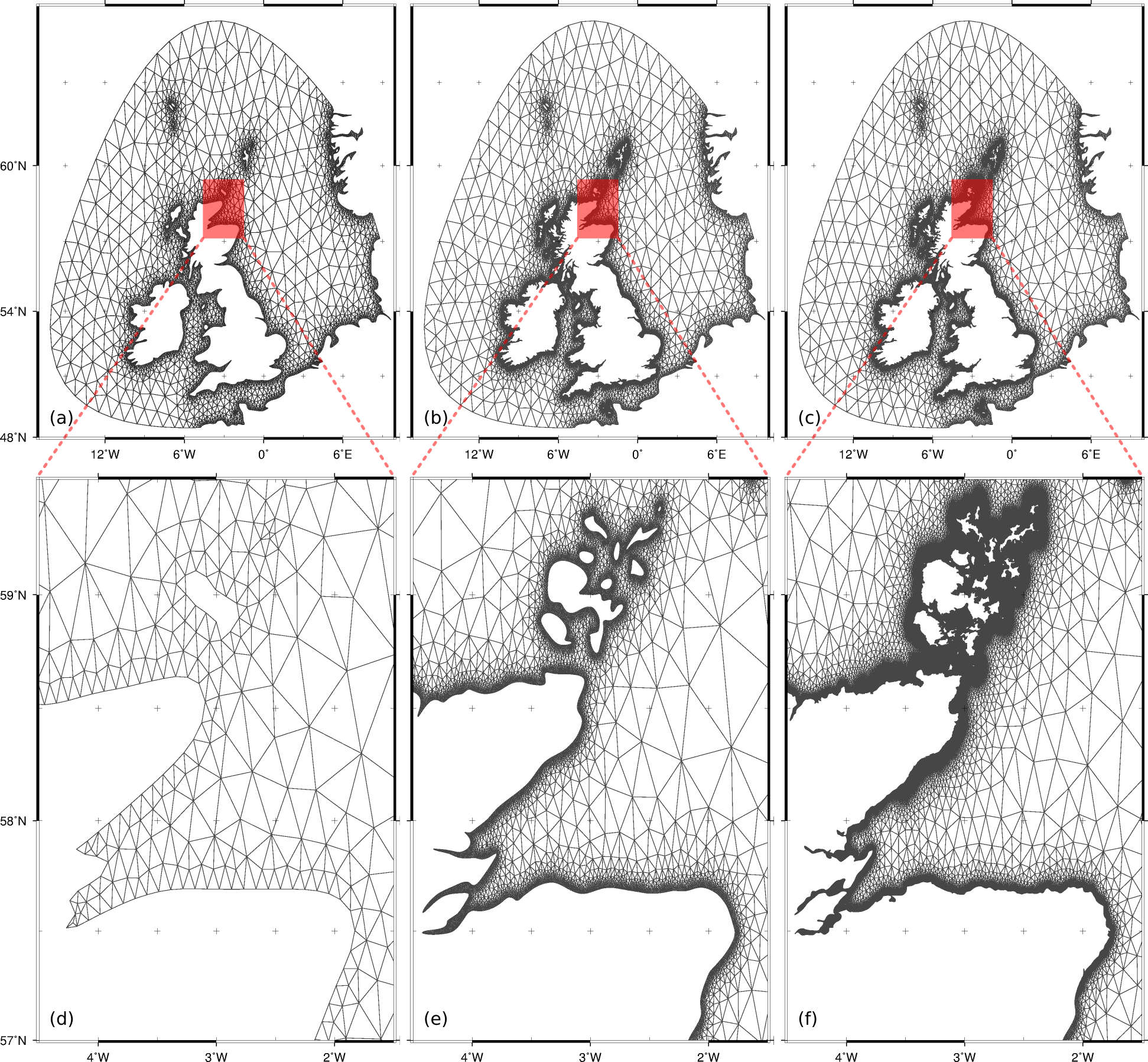}
\caption{Unstructured meshes aimed at tidal resource assessment of coastal regions around the UK, with particular
emphasis on the Orkney Islands and the Severn Estuary. The meshes were constructed from the domain boundaries shown in
figure \ref{fig:simulationDomainMaps}. (a) Coarse mesh composed of $4,307$ vertices, $6,999$ triangles. (b) Intermediate
mesh, composed of $75,368$ vertices, $134,072$ triangles. (c) Fine mesh, composed of $3,382,886$ vertices, $6,696,669$
triangles. (d),(e),(f) Indicated detail, in the region of the Orkney Islands. All meshes were generated in
EPSG:4326 reference system, but are drawn using the Mercator projection.}
\label{fig:simulationDomainMeshes}
\end{figure*}

Figure \ref{fig:simulationDomainMeshes} shows the coarse (panels a and d), intermediate (panels b and e) and
fine unstructured triangular mesh (panels c and f) generated from the domain boundaries in figure \ref{fig:simulationDomainMaps}. All
three meshes are generated in EPSG:4326 (the coordinate reference system axes are longitude and latitude
in degrees). The element size is prescribed in terms of a target edge length. The bathymetry simplification could
be used to calculate a metric based on bathymetry gradient, such that finer resolution is also focused in regions
of steep bathymetry. However, the regions of interest here are relatively close to
the shorelines, with relatively small slopes. Thus the optimal mesh size can be expressed in terms of proximity
functions from shorelines of interest, where detailed reconstructions or full shoreline data are used in the first
place. The maximum edge length is $1.5^\circ$ for all meshes, with angles
measured along a great circle. Different mesh size gradations are used towards the various shoreline reconstructions:
The edge length
at shorelines reconstructed using one mode and $500$ point partition size (violet lines in figure
\ref{fig:simulationDomainMaps}) was $0.1^\circ$, gradating linearly from the shoreline; $0.01^\circ$
at shorelines reconstructed using one mode
and $100$ point partition size (red lines in figure \ref{fig:simulationDomainMaps}), maintained at that size
$0.02^\circ$ from the nearest shoreline; $0.005^\circ$ at shorelines reconstructed using five modes
and $100$ point partition size (blue lines in figure \ref{fig:simulationDomainMaps}), maintained at that size
$0.02^\circ$ from the nearest shoreline;  $0.0005^\circ$ at shorelines reconstructed using the GSHHG full
resolution shoreline (black lines in figure \ref{fig:simulationDomainMaps}), maintained at that size
$0.05^\circ$ from the nearest shoreline. In all gradations, the mesh size increases linearly from the minimum
edge length to the maximum, across a distance of $1^\circ$. The effect of different gradations on mesh
size in the region of the Orkney Islands are shown in Panels (d), (e) and (f) of figure \ref{fig:simulationDomainMeshes}.
The prescribed edge lengths translate to an approximate length of $10$km in the coarse mesh, 
$1$km in the intermediate mesh and $50$ metres in the fine mesh.
The meshes were produced using Gmsh \cite{Geuzaine2009-dd}, by translating the domain boundary and
element size data to formats native to the mesh generator.

\section{Discussion and Conclusion}
A new shoreline and bathymetry simplification method has been introduced,
and results from its application on the shorelines and bathymetry of the NW European continental shelf
have been presented. Existing simplification methods are often based on geometric criteria.
The algorithm presented here is based on principal component analysis, so that the shoreline
or bathymetry can be expressed in terms of a set of modes and corresponding eigen--vectors. A particularly
useful feature of this method is that the modes are calculated such that the most significant structures in the
data can be approximated using just a few modes. Bathymetry and shorelines are examples of multi--scale geometries
where the length--scales of structures extend over multiple orders of magnitude.
Thus a partial reconstruction, using just a few modes instead of all,
will give a smooth approximation of the input shoreline or bathymetry, while capturing the most significant structures.
The results presented in this paper show that the proposed simplification method can perform effective simplification, and the
implementation allows control over the range of scales maintained in the simplified output. Further results
are focused towards mesh generation for tidal flow simulations in
the context of renewable energy generation. High predictive accuracy in such simulations requires meshes with typically very
small element size in the region of interest, so that the smallest scales in the domain geometry are resolved.
Yet in areas further away from the region of interest a larger element size is preferable, in order to reduce
the computational cost, and a simplified geometry is therefore desired. The results presented here demonstrate how the developed simplification
method can be used in this context.
\par
The provision of a framework for algorithmic, or even ad--hoc, processing of
geographical data is one of the targets of Geographical Information
Systems (GIS). In the context of GIS, data is broadly
classified into vector and raster data--structures, and the implementation of the method reflects this
classification. In this way, the implementation allows for simplification of any vector or raster
data--sets, albeit the focus here is on shorelines and bathymetry.
\par
Ongoing work is aimed at extending bathymetry processing towards combinations of very high resolution data--sets
over a small area with lower--resolution data over a wider area. For example, blending a high resolution, high accuracy
bathymetric survey of a region earmarked for tidal turbine installations with lower resolution and less accurate
bathymetry data over the rest of the simulation domain. Hydrodynamic simulations on more elaborate
meshes, where a minimum element edge length is chosen so that power--extracting devices and infrastructure
can be resolved or parameterised are also underway, with mesh formats being designed to work with the Fluidity,
OpenTidalFarm, Telemac and MIKE models.
In terms of the implementation, future work will focus on improvements such as automatic detection
of intersecting shorelines, parallelisation, as well as releasing the software source code under a
permissive open--source licence.

\section*{Acknowledgments}
This work was supported by an EPSRC Impact Acceleration Award (EP/K503733/1)
and EPSRC grants EP/J010065/1, EP/M011054/1.  The authors would also like to 
acknowledge the support of the Imperial College High Performance Computing Service.
%


\bibliographystyle{IEEEtran}
\bibliography{EWTEC2015PaperBibliography.bib}

\begin{thebibliography}{10}
\providecommand{\url}[1]{#1}
\csname url@samestyle\endcsname
\providecommand{\newblock}{\relax}
\providecommand{\bibinfo}[2]{#2}
\providecommand{\BIBentrySTDinterwordspacing}{\spaceskip=0pt\relax}
\providecommand{\BIBentryALTinterwordstretchfactor}{4}
\providecommand{\BIBentryALTinterwordspacing}{\spaceskip=\fontdimen2\font plus
\BIBentryALTinterwordstretchfactor\fontdimen3\font minus
  \fontdimen4\font\relax}
\providecommand{\BIBforeignlanguage}[2]{{%
\expandafter\ifx\csname l@#1\endcsname\relax
\typeout{** WARNING: IEEEtran.bst: No hyphenation pattern has been}%
\typeout{** loaded for the language `#1'. Using the pattern for}%
\typeout{** the default language instead.}%
\else
\language=\csname l@#1\endcsname
\fi
#2}}
\providecommand{\BIBdecl}{\relax}
\BIBdecl

\bibitem{Hasegawa2011}
D.~Hasegawa, J.~Sheng, D.~A. Greenberg, and K.~R. Thompson, ``{Far-field
  effects of tidal energy extraction in the Minas Passage on tidal circulation
  in the Bay of Fundy and Gulf of Maine using a nested-grid coastal circulation
  model},'' \emph{Ocean Dynamics}, vol.~61, no.~11, pp. 1845--1868, 2011.

\bibitem{Funke2014-vk}
S.~W. Funke, P.~E. Farrell, and M.~D. Piggott, ``{Tidal turbine array
  optimisation using the adjoint approach},'' \emph{Renewable Energy}, vol.~63,
  pp. 658--673, 2014.

\bibitem{Martin-Short2015}
R.~Martin-Short, J.~Hill, S.~C. Kramer, A.~Avdis, P.~A. Allison, and M.~D.
  Piggott, ``{Tidal resource extraction in the Pentland Firth, UK: potential
  impacts on flow regime and sediment transport in the Inner Sound of
  Stroma},'' \emph{{Renewable Energy}}, vol.~76, pp. 596--607, 2015.

\bibitem{gorman:2007}
\BIBentryALTinterwordspacing
G.~J. Gorman, M.~D. Piggott, and C.~C. Pain, ``Shoreline approximation for
  unstructured mesh generation,'' \emph{Computers and Geosciences}, vol.~33,
  no.~5, pp. 666--677, 2007. [Online]. Available:
  \url{http://www.sciencedirect.com/science/article/pii/S0098300406002020}
\BIBentrySTDinterwordspacing

\bibitem{gorman:2008}
\BIBentryALTinterwordspacing
G.~J. Gorman, M.~D. Piggott, M.~R. Wells, C.~C. Pain, and P.~A. Allison, ``{A
  systematic approach to unstructured mesh generation for ocean modelling using
  GMT and Terreno},'' \emph{Computers and Geosciences}, vol.~34, no.~12, pp.
  1721 -- 1731, 2008. [Online]. Available:
  \url{http://www.sciencedirect.com/science/article/pii/S0098300408001003}
\BIBentrySTDinterwordspacing

\bibitem{Ramer:1972}
\BIBentryALTinterwordspacing
U.~Ramer, ``{An iterative procedure for the polygonal approximation of plane
  curves},'' \emph{Computer Graphics and Image Processing}, vol.~1, no.~3, pp.
  244--256, 1972. [Online]. Available:
  \url{http://dx.doi.org/10.1016/S0146-664X(72)80017-0}
\BIBentrySTDinterwordspacing

\bibitem{douglas_peucker:1973}
D.~Douglas and T.~Peucker, ``Algorithms for the reduction of the number of
  points required to represent a digitized line or its caricature,'' \emph{The
  Canadian Cartographer}, vol.~10, no.~2, pp. 112--122, 1973.

\bibitem{wessel_smith:1996}
\BIBentryALTinterwordspacing
P.~Wessel and W.~H.~F. Smith, ``A global, self-consistent, hierarchical,
  high-resolution shoreline database,'' \emph{Journal of Geophysical Research:
  Solid Earth}, vol. 101, no.~B4, pp. 8741--8743, 1996. [Online]. Available:
  \url{http://dx.doi.org/10.1029/96JB00104}
\BIBentrySTDinterwordspacing

\bibitem{Demsar_et_al:2013}
U.~Dem\v{s}ar, P.~Harris, C.~Brunsdon, S.~A. Fotheringham, and S.~McLoone,
  ``{Principal Component Analysis on Spatial Data: An Overview},'' \emph{Annals
  of the Association of American Geographers}, vol. 103, no.~1, pp. 106--128,
  2013.

\bibitem{ruessink_et_al:2004}
\BIBentryALTinterwordspacing
B.~G. Ruessink, I.~M.~J. van Enckevort, and Y.~Kuriyama, ``{Non--linear
  principal component analysis of nearshore bathymetry},'' \emph{Marine
  Geology}, vol. 203, no. 1--2, pp. 185 -- 197, 2004. [Online]. Available:
  \url{http://www.sciencedirect.com/science/article/pii/S0025322703003347}
\BIBentrySTDinterwordspacing

\bibitem{medina_et_al:1992}
R.~Medina, C.~Vidal, M.~Losada, and A.~Roldan, ``{Three--mode principle
  component analysis of bathymetric data, applied to ``Playa de Castilla''
  (Huelva, Spain)},'' \emph{Coastal Engineering Proceedings}, vol.~1, no.~23,
  1992.

\bibitem{winant_et_al:1975}
\BIBentryALTinterwordspacing
C.~D. Winant, D.~L. Inman, and C.~E. Nordstrom, ``Description of seasonal beach
  changes using empirical eigenfunctions,'' \emph{Journal of Geophysical
  Research}, vol.~80, no.~15, pp. 1979--1986, 1975. [Online]. Available:
  \url{http://dx.doi.org/10.1029/JC080i015p01979}
\BIBentrySTDinterwordspacing

\bibitem{galton:1889}
F.~Galton, \emph{Natural Inheritance}.\hskip 1em plus 0.5em minus 0.4em\relax
  London: Macmillan, 1889.

\bibitem{pearson:1901}
\BIBentryALTinterwordspacing
K.~Pearson, ``{LIII. On lines and planes of closest fit to systems of points in
  space},'' \emph{Philosophical Magazine Series 6}, vol.~2, no.~11, pp.
  559--572, 1901. [Online]. Available:
  \url{http://dx.doi.org/10.1080/14786440109462720}
\BIBentrySTDinterwordspacing

\bibitem{hotelling:1933}
\BIBentryALTinterwordspacing
H.~Hotelling, ``Analysis of a complex of statistical variables into principal
  components,'' \emph{Journal of Educational Psychology}, vol.~24, no.~6, pp.
  417--441, 1933. [Online]. Available:
  \url{http://psycnet.apa.org/doi/10.1037/h0071325}
\BIBentrySTDinterwordspacing

\bibitem{hotelling:1936}
------, ``Relationships between two sets of variates,'' \emph{Biometrika},
  vol.~28, pp. 321--377, 1936.

\bibitem{rosenfeld_kak:1982}
{Digital Picture Processing}, \emph{Rosenfeld, Azriel and Kak, Avinash
  C}.\hskip 1em plus 0.5em minus 0.4em\relax Academic Press, 1982.

\bibitem{holmes_lumley_berkooz:1996}
P.~Holmes, J.~L. Lumley, and G.~Berkooz, \emph{{Turbulence, Coherent
  Structures, Dynamical Systems and Symmetry}}.\hskip 1em plus 0.5em minus
  0.4em\relax Cambridge University Press, 1996.

\bibitem{jolliffe_pca_book:2002}
I.~T. Jolliffe, \emph{Principal Component Analysis}.\hskip 1em plus 0.5em minus
  0.4em\relax Springer, 2002.

\bibitem{wold_et_al:1987}
\BIBentryALTinterwordspacing
S.~Wold, K.~Esbensen, and P.~Geladi, ``Principal component analysis,''
  \emph{{Chemometrics and Intelligent Laboratory Systems}}, vol.~2, no. 1--3,
  pp. 37--52, 1987, {Proceedings of the Multivariate Statistical Workshop for
  Geologists and Geochemists}. [Online]. Available:
  \url{http://www.sciencedirect.com/science/article/pii/0169743987800849}
\BIBentrySTDinterwordspacing

\bibitem{willis_et_al:2010}
\BIBentryALTinterwordspacing
M.~Willis, I.~Masters, S.~Thomas, R.~Gallie, J.~Loman, A.~Cook, R.~Ahmadian,
  R.~Falconer, B.~Lin, G.~Gao, M.~Cross, N.~Croft, A.~Williams, M.~Muhasilovic,
  H.~Ian, R.~Fidler, C.~Wooldridge, I.~Fryett, P.~Evans, T.~O’Doherty,
  D.~O’Doherty, and A.~Mason-Jones, ``{Tidal turbine deployment in the
  Bristol Channel: a case study},'' \emph{Proceedings of the Institute of Civil
  Engineers--Energy}, vol. 163, no.~3, pp. 93--105, 2010. [Online]. Available:
  \url{http://www.icevirtuallibrary.com/content/article/10.1680/ener.2010.163.3.93}
\BIBentrySTDinterwordspacing

\bibitem{Geuzaine2009-dd}
C.~Geuzaine and J.-F. Remacle, ``{Gmsh: A 3--D finite element mesh generator
  with built-in pre- and post-processing facilities},'' \emph{Int. J. Numer.
  Methods Eng.}, vol.~79, no.~11, pp. 1309--1331, 2009.

\end{thebibliography}
%
%

\end{document}